\documentstyle[psfig]{l-school}
\begin{document}
\markboth{Uwe Trittmann}{POSITRONIUM: MULTIPLETS}
\setcounter{part}{13}
%
\title{Front form QED(3+1): The spin-multiplet structure of the
             positronium spectrum at strong coupling\footnote{
Invited talk at the International Workshop
``New Non-Perturbative Methods and Quantization on the Light Cone'', 
  Les Houches (France), February 24-March 7, 1997;
  to appear in the proceedings.}}
\author{U. Trittmann}
\institute{Max--Planck-Institut f\"ur\\ Kernphysik\\
Postfach 10 39 80, D--69029 Heidelberg, \\
Germany}

\maketitle
%
%
\vspace*{-2cm}
\section{INTRODUCTION}

The practitioner  has to think about (at least) two obstacles, when he wants to 
solve a Hamiltonian field theory problem in {\em front form} dynamics. 
One is the problem that the theory is not manifestly rotational invariant.
The other is the question, how one should construct effective theories at all
in this framework.
Of course, the first problem exists analogously in instant form dynamics,
where the theory is not manifestly Lorentz invariant. Connected to this
problem is the fact that
any symmetry which is not manifest will break down immediately with 
the slightest approximation in a Hamiltonian field theory.
There is broad agreement that the construction of  
effective theories is inevitable, if one wants to describe the
low energy region, {\em i.e.}~the bound states, of a strongly coupled theory. 
A variety of formalisms were suggested concerning this topic 
\cite{Wilson94}\cite{Wegner94}\cite{Pauli96b}.

\begin{figure}
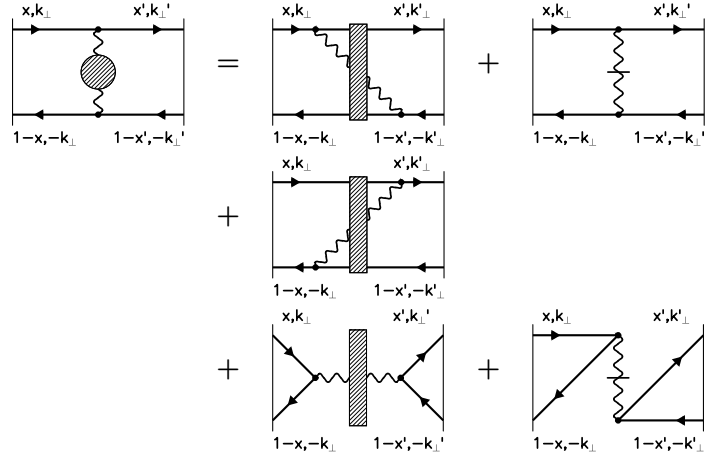

\centerline{
\begin{tabular}{ccccccc}
\psfig{figure=all_effective.epsi,width=2.3cm,angle=-90}
&\hspace{0.0cm}\raisebox{1.0cm}[-1.4cm]{\bf =}&  
\psfig{figure=dyn_effective.epsi,width=2.3cm,angle=-90} 
&\raisebox{1.0cm}[-1.4cm]{\bf +} 
&\psfig{figure=seagull.epsi,width=2.3cm,angle=-90} \\
&\raisebox{1.0cm}[-1.4cm]{\bf +}&
\psfig{figure=dyn2_effective.epsi,width=2.3cm,angle=-90}
& &\\
&\raisebox{1.0cm}[-1.4cm]{\bf +}&
\psfig{figure=anni_effective.epsi,width=2.3cm,angle=-90}
&\raisebox{1.0cm}[-1.4cm]{\bf +}
& \psfig{figure=anni_seag.epsi,width=2.3cm,angle=-90}
\end{tabular}}
\caption{\label{PosiEff}
The graphs of the positronium model. Effective photon lines are 
labeled by hashed rectangles. The graphs of the annihilation interaction are
at the bottom line.}
\end{figure}

In this article I shall present quantitative investigations concerning
the above two problems with a specified model, namely 
positronium as a  QED(3+1) bound state.
The model is represented by the energy diagrams of Fig.~\ref{PosiEff}.
To be more specific, an effective theory was constructed following the work 
of {\sc Pauli}\cite{Pauli96b}. The effective matrix elements were calculated
analytically,
and the emerging integral equation was put on the computer to numerically 
solve for the eigenvalues and eigenfunctions of the Hamiltonian.

Proceeding in this manner, one can answer the following 
questions {\em quantitatively}:
Are the results of front form theory rotationally invariant, {\em i.e.}
do states of different $J_z$ form multiplets?
How good is the underlying effective theory concerning its cutoff dependence?
Is the multiplet structure of the spectrum reproduced correctly? 
And: what is the numerical evidence for the degeneracy of corresponding states?

\section{METHOD} 

A complete description of the applied method is surely beyond the 
scope of this article. I review shortly its major steps.

To get a finite dimensional Hamiltonian out of the infinite dimensional
canonical Hamiltonian, we introduce a cutoff $\Lambda$ on the kinetic energies.
We then (formally) map an $N\times N$ onto a $2 \times 2$ block matrix
by subsequent projections (Bloch-Feshbach formalism) and call these sectors 
$P$-{\em space} and $Q$-{\em space}.  
Note that the $Q$-space is already an effective sector by construction:
\[
H_{Q}=H_{Q}^{(N)}+X^{\rm eff}_{Q},
\]
 where $H_{Q}^{(N)}$ is a piece of
the original Hamiltonian and $X^{\rm eff}_{Q}$ are the interactions 
generated by the projections. 
The Hamiltonian reads
\[
H_{\rm LC} = \left(
                \begin{array}{cc}
                        H_{ P} & H_{ PQ}\\
                        H_{ QP} & {H}_{ Q}
                \end{array}
             \right). 
\]
The final projection maps the $Q$-space onto the $P$-space.
One solves for the state $\hat{Q}|\psi\rangle = |e\bar{e}\gamma\rangle$ 
with a resolvent involving the {\em redundant parameter} $\omega$ 
\[
G(\omega):=\langle Q|\omega - H_{\rm LC}|Q\rangle ^{-1},
\]
and obtains the {\em nonlinear} equation
\[
H^{\rm eff}_{\rm LC}(\omega) |\psi_n(\omega)\rangle = 
M_n^2(\omega) |\psi_n(\omega)\rangle, 
\]
with the effective Hamiltonian ($\hat{P}|\psi\rangle = |e\bar{e}\rangle$)
\[
H^{\rm eff}_{\rm LC}(\omega) := \hat{P}H_{\rm LC}\hat{P} +  
 \hat{P}H_{\rm LC}\hat{Q}(\omega - H_{\rm LC})^{-1} \hat{Q}H_{\rm LC}\hat{P}.
\]
The fixing of the redundant parameter works as follows.
We divide the sector Hamiltonian in $Q$-space 
into $H_{Q}=M^2_Q+V_Q$,
and the interaction into a diagonal part
$\langle V_Q\rangle$ and a non-diagonal part $\delta V_Q$
\[
V_Q=\langle V_Q\rangle+\delta V_Q.
\]
With the definition
$ 
T^* := \omega-\langle V_Q\rangle
$
we expand the resolvent around the diagonal interaction 
$\langle V_Q\rangle$ 
\begin{eqnarray}
\frac{1}{\omega-H_Q}&=&\frac{1}{T^*-M^2_Q}+\frac{1}{T^*-M^2_Q}
\delta V_Q\frac{1}{T^*-M^2_Q-\delta V_Q}\nonumber\\
&\simeq&\frac{1}{T^*-M^2_Q}.\label{approximation}
\end{eqnarray}
The approximation in the last step is to consider the first term only. 
It has severe 
consequences: a collinear singularity occurs, 
proportional to
\[
\frac{1}{\cal D}\frac{\Delta(x,k_{\perp},x',k_{\perp}';T^*)}{|x-x'|},\quad
\mbox{}
\quad{with}\quad {\cal D}(x,x';T^*):=|x-x'|(T^*-M^2_Q).
\]
We can calculate $\Delta(x,k_{\perp},x',k_{\perp}';T^*)$ with $P$-space graphs
\[
\Delta=M^2_{Q}-\omega-\frac{(k_{\perp}-k_{\perp}')^2}{x-x'}
-\frac{1}{2}\left(l^-_e-l^-_{\bar{e}}\right),
\]
where
$l_e^{\mu}:=(k_e'-k_e)^{\mu}$ and
$l_{\bar{e}}^{\mu}:=(k_{\bar{e}}-k_{\bar{e}}')^{\mu}$ are the momentum
transfers.
One determines the parameter
$\omega$ by demanding that the {\em collinear} singularity, induced by the 
approximation, Eq.~(\ref{approximation}),
vanishes
\[
\Delta(x,k_{\perp},x',k_{\perp}';T^*)=0, \quad\quad 
\forall x,x',k_{\perp},k_{\perp}'. 
\]
This allows for the calculation of an explicit expression for $T^*$
\[
T^*(x,\vec{k}_{\perp};x',\vec{k}'_{\perp})
        =\frac{1}{2}\left( \frac{m_f^2 + 
        \vec{k}_{\perp}}{x(1-x)} + \frac{m_f^2 + \vec{k}'_{\perp}}{x'(1-x')}
        \right).
\]
The interpretation is that $T^*$ is an approximation of the summed
interactions of the higher Fock states.

Finally, the procedure yields an integral equation.
The effective Hamiltonian operates only in $P$-space, and 
the continuum version of the eigenvalue problem is
\begin{eqnarray*}
0&=&\left(\frac{m_f^2 + \vec{k}^2_{\perp}}{x(1-x)}- M_n^2\right)
\psi_n(x,\vec{k}_{\perp};\lambda_1,\lambda_2)\\
&&\frac{g^2}{16\pi^3}
\sum_{\lambda'_1,\lambda'_2}\int_D \frac{dx'd^2\vec{k}'_{\perp}}
{\sqrt{xx'(1-x)(1-x')}}
\frac{j^{\mu}(l_e,\lambda_e)j_{\mu}(l_{\bar{e}},\lambda_{\bar{e}})}
{l_e^{\mu}l_{e,\mu}}\\
&&\times \psi_n(x',\vec{k}'_{\perp};\lambda'_1,\lambda'_2),
\end{eqnarray*}
with an integration domain $D$ defined by a cutoff on the kinetic energy
of the states.
One observes that the effective interaction
is gauge invariant and a Lorentz scalar
\[
U_{\rm eff}:=\frac{j^{\mu}(l_e,\lambda_e)j_{\mu}(l_{\bar{e}},\lambda_{\bar{e}})}
{l_e^{\mu}l_{e,\mu}}.
\]

\begin{figure}[h]
\centerline{\psfig{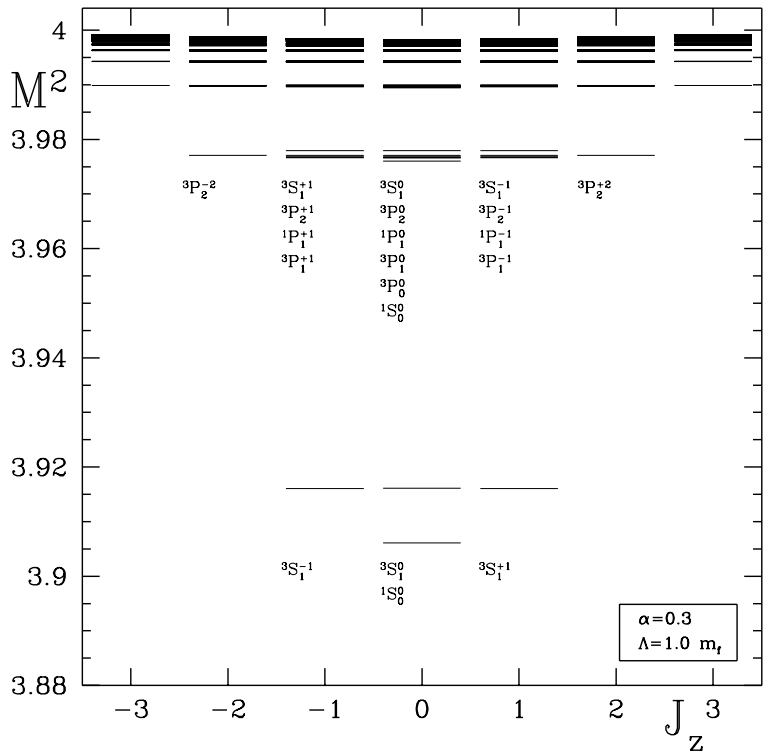}}
\caption{Positronium spectrum in different $J_z$ sectors. $N_1=N_2=21$.}
\label{yrast}
\end{figure}
\section{RESULTS}

As pointed out in the last section, the goal is to construct an effective
Hamiltonian describing the positronium bound states.
We showed how to find a reasonable {\em ansatz} for contributions of higher 
Fock states. To find out about the rotational symmetry of the theory, 
one has to observe that the rotation operator ${\cal J}_3$ around the $z$ axis 
is kinematic. Consequently, states can be classified according to the 
corresponding quantum number $J_z$. We have shown in 
Refs.~\cite{TrittmannPauli97a}\cite{TrittmannPauli97b} that closed expressions
for the matrix elements
of the {\em effective Hamiltonian} can be calculated, even for arbitrary $J_z$.
This enables one to solve for the positronium eigenvalues in each sector 
of $J_z$ separately and to compare the results to find out about possible
degeneracies.
A computer code for solving these eigenvalue problems,
{\em i.e.}~{\em integral equations}, has been generated with correct 
{\em Coulomb counterterms}. The latter is important to guarantee 
numerical stability and precision high enough to be able to give 
a quantitative statement concerning the (non)degeneracy of states.
In fact, the convergence of the eigenvalues turns out to be exponential in
the number of integration points. It is plotted in Fig.~\ref{specanni}.

\begin{figure}[h]
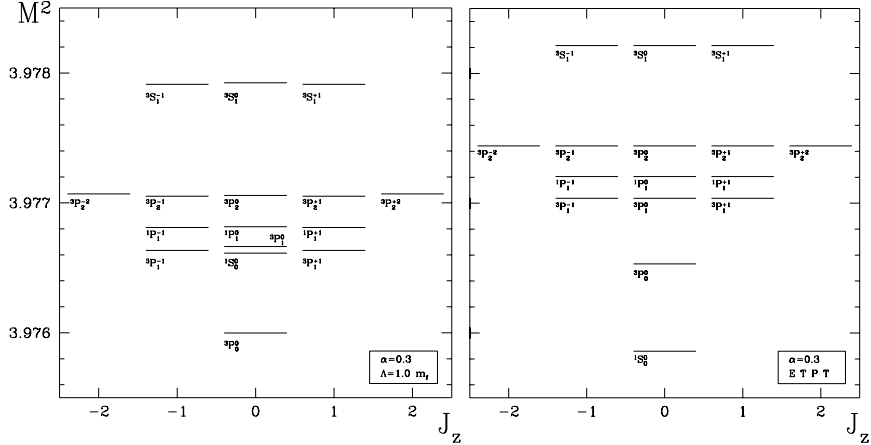

\centerline{
\psfig{figure=yrast_n2.epsi,width=6cm,angle=0}
\psfig{figure=yrast_n2_E.epsi,width=5.31cm,angle=0}}
\caption{\label{yrastn2}Comparison of 
multiplets for $n{=}2$: (a) results of the present work with
$\alpha=0.3$, $\Lambda= 1.0\,m_f$, $N_1=N_2=21$; 
(b) equal-time perturbation theory (ETPT) up to order ${\cal O}(\alpha^4)$.}
\end{figure}

\begin{figure}[t]
\centerline{
\psfig{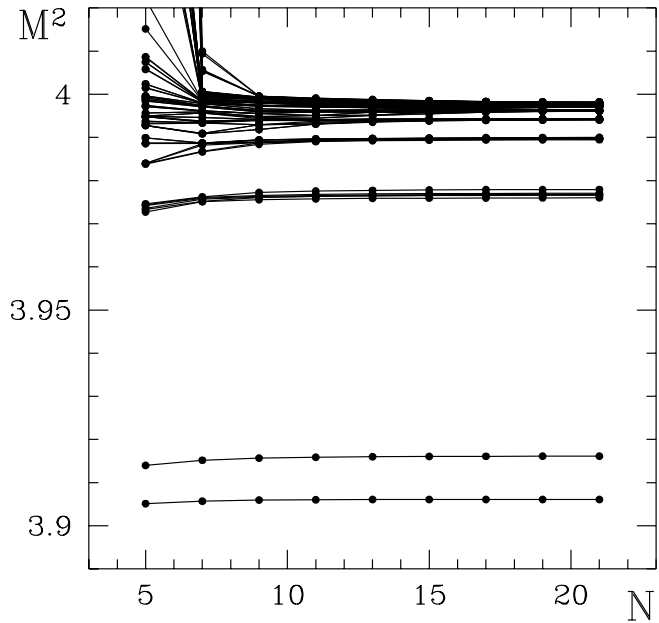}}
\caption
{\protect\label{specanni}The positronium spectrum including the annihilation 
channel in the $J_z{=}0$ sector.  
Parameters of the calculation: $\alpha=0.3, 
\Lambda=1.0\, m_f$. 
The mass squared eigenvalues $M^2_n$ in units of the electron mass $m^2_f$
are shown as functions of the number of integration points $N\equiv N_1=N_2$.
The triplet states, especially $1^3S_1$, are lifted up, the singlet
mass eigenvalues are the same as without the annihilation channel.}
\end{figure}

The results are compared to equal time perturbation theory (ETPT).
The spectrum is shown in Fig.~\ref{yrast}, which is the most important of this
article. A one-to-one comparison between the multiplets for a Bohr quantum 
number $n=2$ is depicted in Fig.~\ref{yrastn2}. 
It is quite clear how to interpret these results.
One reads off easily from the number of degenerate $J_z$ levels the
quantum number $J$ of the complicated
operator $J^2$, consisting of the angular momentum operators, 
{\em cf.}~the theory of the {Poincar\'e group} 
in front form dynamics \cite[Chapter 3.1]{TrittmannPauli97a}.

It is worthwhile mentioning another non-trivial result of the formalism 
considered in this article.
The annihilation channel can be included 
straightforwardly, which is an intrinsic check of the
effective theory.
Surprisingly the instantaneous and the dynamic graphs act in 
different $J_z$ sectors.
An interesting feature concerning the annihilation channel
is the calculation of the hyperfine splitting, where this
channel plays a major role and yields a well-known contribution.
The coefficient of the hyperfine splitting is defined as
\begin{eqnarray*}
C_{hf}&=&\frac{M_{t}-M_{s}}{\alpha^4}=
\frac{1}{2}\left[\frac{2}{3}+\left(\frac{1}{2}\right)-
\frac{\alpha}{\pi}\left(\ln 2 +\frac{16}{9}\right)
+{\cal O}(\alpha^2)\right].
\end{eqnarray*}
The results for the two formalisms are
\[
C_{hf}= \left \{                
\begin{array}{rl}
0.56 \quad &; \mbox{this work}\\
\frac{7}{12} \simeq 0.58 \quad &; \mbox{ETPT[${\cal O}(\alpha^4)$]},
\end{array}
\right.
\]
which are in reasonable agreement.

Let us have a closer look at the results. 
One question listed above was, how good is the 
numerical evidence for degeneracies.
One can plot the deviation of 
corresponding eigenvalues for $J_z{=}0$ and $J_z{=}1$ multiplets
with growing number of integration points, as was done in Fig.~12 of 
Ref.~\cite{TrittmannPauli97b}. 
A $\chi^2$ fit of the difference function to
\[
\Delta M^2(N)=a-b\exp\left\{(N-N_0)/c\right\}.
\]
yields for the ground state triplet $1\,^{3\!}S_1$:
$a= -(5.47\pm 0.95)\times 10^{-5}$, $b= (1.88\pm 0.03)\times 10^{-4}$, 
$c= 4.03\pm 0.11$. It seems thus quite justified to consider these states 
degenerate.

The cutoff dependence of the eigenvalues was investigated, too. For the actual
graphs, see Ref.~\cite{Trittmann97a}. The role of the annihilation channel 
seems to be to stabilize the eigenvalues.
For the triplet ground state we can fit the dependence on $log\; \Lambda$ with
\[
M^2_{t}(\Lambda)=\left\{\begin{array}{ll}
3.90976-0.01858 \log \Lambda+0.00789 \log^2\Lambda &\quad ; \mbox{no annih.}\\
3.91392-0.00029 \log\Lambda+ 0.00015 \log^2\Lambda &\quad ; \mbox{incl.~annih.}
\end{array}
\right.
\]
The decrease of the triplet 
with $\log\Lambda$ is suppressed by the inclusion of the 
annihilation channel by a factor of 60!
Care is, however, to be taken, since the annihilation channel is absent in 
all QED bound states not built out of a particle and its antiparticle 
\cite{Brodsky97}.

\begin{figure}
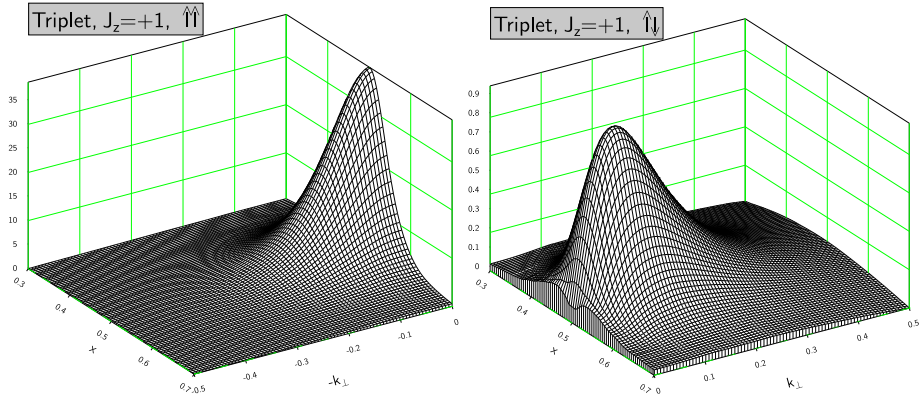

\centerline{
\psfig{figure=wf_0j10.epsi,width=6cm,angle=0}
\psfig{figure=wf_0j11.epsi,width=6cm,angle=0}
}
\caption{
The triplet 
ground state wavefunction for $J_z{=}+1$ as a function of 
the longitudinal momentum fraction $x$ and the transverse momentum $k_{\perp}$,
omitting the dependence on the angle $\varphi$. The
calculation was done with $\alpha=0.3,\Lambda=1.0\, m_f, N_1=41, N_2=11$.
Shown are: (a) ($\uparrow\uparrow$)-component, 
(b) ($\uparrow\downarrow$)-component.}
\label{wfexcited}
\end{figure}

\section{CONCLUSIONS}

To conclude, we review the results of the positronium theory
in {\em front form} dynamics as presented in this article.

The correct positronium spectrum is obtained in all $J_z$ sectors. 
The $J_z$ states form (degenerate) multiplets, which means that 
rotational invariance is restored in the solution of a Hamiltonian 
field theory in {\em front form} dynamics.
Consequently, there is no need to diagonalize the complicated 
rotation operator ${\cal J}^2$, because its eigenvalues can be read off 
by counting the number of degenerate eigenstates within the $J_z$ multiplets.
The annihilation channel can be introduced into the theory straightforwardly.
Furthermore, the  annihilation channel seems important for the cutoff behavior.
From these facts, we can deduce 
that the applied underlying effective theory is correct.
The computer code which was generated to solve for the eigenstates is 
applicable also to other effective Hamiltonians. The application to 
{\sc Wegner's} formalism \cite{Wegner94} is work in progress.
As an outlook one could think of plugging a running coupling constant
into the code and to calculate the meson spectrum. Surely,  this implies 
a deeper
understanding of the numerical behavior of the occuring severe singularities.

\ack{It is a pleasure to 
thank the organizers of the workshop for the invitation, hospitality 
and, last but not least, for the financial support.}
%

%
%
%
%
\end{document}